\documentclass[prb,showpacs,superscriptaddress,onecolumn,amssymb,amsfonts]{revtex4}
\usepackage{amsmath}
\usepackage{amssymb}
\usepackage{bm}
\usepackage{epsfig}
\usepackage{graphics}
\usepackage{dsfont}
\def\prd{Phys. Rev. D}
\def\prb{Phys. Rev. B}
\def\beq{\begin{equation}}
\def\eeq{\end{equation}}
\def\beqn{\begin{eqnarray}}
\def\eeqn{\end{eqnarray}}

\renewcommand{\bf}{\mathbf}

\begin{document}
\title{Topological $BF$ field theory description of topological insulators}
\author{Gil Young Cho}
\affiliation{Department of Physics, University of California,
Berkeley, CA 94720}
\author{Joel E. Moore}
\affiliation{Department of Physics, University of California,
Berkeley, CA 94720} \affiliation{Materials Sciences Division,
Lawrence Berkeley National Laboratory, Berkeley, CA 94720}
\date{\today}


%

\begin{abstract}
Topological phases of matter are described universally by topological field theories in the same way that symmetry-breaking phases of matter are described by Landau-Ginzburg field theories.  We propose that topological insulators in two and three dimensions are described by a version of abelian $BF$ theory.  For the two-dimensional topological insulator or quantum spin Hall state, this description is essentially equivalent to a pair of Chern-Simons theories, consistent with the realization of this phase as paired integer quantum Hall effect states.  The $BF$ description can be motivated from the local excitations produced when a $\pi$ flux is threaded through this state.  For the three-dimensional topological insulator, the $BF$ description is less obvious but quite versatile: it contains a gapless surface Dirac fermion when time-reversal-symmetry is preserved and yields ``axion electrodynamics'', i.e., an electromagnetic $E \cdot B$ term, when time-reversal symmetry is broken and the surfaces are gapped.  Just as changing the coefficients and charges of 2D Chern-Simons theory allows one to obtain fractional quantum Hall states starting from integer states, $BF$ theory could also describe (at a macroscopic level) fractional 3D topological insulators with fractional statistics of point-like and line-like objects.
\end{abstract}

\maketitle

\section{Introduction}

Most ordered phases of condensed matter can be classified using the concept of symmetry breaking, which leads to a theoretical description in terms of Landau-Ginzburg field theories.  Such theories have had remarkable successes, such as the quantitative explanation of critical phenomena.  Topological phases such as the quantum Hall effect require a different type of description in terms of ``topological field theories''.  These theories have the remarkable property that they are independent of the spacetime metric.  The most famous example is the description of quantum Hall phases by Chern-Simons theory, which we review in a moment.  As shown by Wen, that theory leads to detailed predictions about edge tunneling experiments on fractional quantum Hall samples, which to some degree have been verified at least qualitatively.

Recently new topological phases of electrons with time-reversal symmetry have been discovered.  These ``topological insulators''~\cite{hasankane,moorenature} result from spin-orbit coupling and exist both in two~\cite{kane&mele-2005,zhangscience1,molenkampscience} and three~\cite{fu&kane&mele-2007,moore&balents-2006,rroy3D,hsieh} dimensions, unlike the quantum Hall effect, which is essentially two-dimensional.  These phases are insulating in bulk but support conducting edge or surface states.  Hence one picture of a topological insulator is as a bulk insulator that necessarily has conducting edge or surface states; more recently there has been considerable interest in the electromagnetic response~\cite{qilong,essinmoorevanderbilt} when a weak time-reversal-breaking perturbation gaps the surface state of the three-dimensional (3D) topological insulator.   Our goal in this work is to find an effective field theory description of topological insulator phases in the same spirit as the Chern-Simons effective theory of the quantum Hall effect.

In order to understand why we do not consider the electromagnetic response in the 3D topological insulator to be a topological field theory in the sense of Chern-Simons theory, it is useful to review the quantum Hall case.  In the quantum Hall effect, the Chern-Simons theory appears in two forms.  We assume for now a single-component theory, such as describes the Laughlin states at $\nu = 1/k$.  In terms of an internal gauge field $a_\mu$ and an integer $k$, the Chern-Simons Lagrangian density is
\beq
L_{CS} = \frac{k}{4\pi}\varepsilon^{\mu\nu\lambda}a_{\mu}\partial_{\nu}a_{\lambda}.
\label{csl}
\eeq
Upon coupling this term to electromagnetic fields $A_\mu$ by adding the term
\beq
L_c = - J^\mu A_\mu, \quad J^\mu = {1 \over 2 \pi} \epsilon^{\mu \nu \lambda} \partial_\nu a_\lambda
\eeq
integrating out the internal gauge field $a_\mu$ generates a term for the electromagnetic gauge field $A_\mu$ that has the {\it same} Chern-Simons form as the internal gauge theory, but with coefficient $-{1 \over 4 \pi k}$ rather than ${k \over 4 \pi}$.  (Note that $A_\mu$ does not include the background magnetic field generating the quantum Hall state.)

So for the quantum Hall effect, both the internal theory and the electromagnetic term have the same Chern-Simons form.  The Chern-Simons theory on a closed two-dimensional manifold has a set of $k^g$ zero-energy states, where $g$ is the genus of the manifold, and an infinite gap to other states.  The quantum Hall response is generated by the ground state's response to electromagnetic perturbations in similar fashion to the superflow in a superconductor.  On a manifold with boundary, the bulk Chern-Simons theory alone is not gauge-invariant, because under a gauge transformation, (\ref{csl}) changes by a total derivative.  This total derivative leads to the topological part of the gapless edge theory; combining it with non-universal interactions from the edge gives the chiral Luttinger liquid description of edge states.

Our general goal is to carry through a similar program for topological insulators.  We note that many features of three-dimensional topological insulators with gapped surfaces have been understood in terms of a total-derivative term for electromagnetic fields known as ``axion electrodynamics''~\cite{wilczekaxion,qilong}:
\beq
{\cal L}_{EM} = \frac{\theta e^2}{2\pi h} {\bf E} \cdot {\bf B} =
\frac{\theta e^2}{16\pi h} \epsilon^{\alpha \beta \gamma \delta}
	F_{\alpha \beta} F_{\gamma \delta}.
\label{axioncoupling}
\eeq
This is analogous to the electromagnetic Chern-Simons term for $A^\mu$ in the quantum Hall effect.  It is ``topological'' in two senses: it is a total derivative of the electromagnetic Chern-Simons term, indicating that the surfaces where $\theta$ changes support a half-integer quantum Hall effect.  Its microscopic origin in a crystal is definitely topological as it involves the Chern-Simons form of the Berry connection of the Bloch electrons~\cite{qilong,essinmoorevanderbilt,essinturnermoorevanderbilt,malashevich}.

We will argue that the natural description of the internal gauge theory takes a different form, which is one important difference between the quantum Hall and topological insulator cases.  Note that one difference between the axion coupling \eqref{axioncoupling} and the Chern-Simons term is that the axion term is gauge-invariant, while the Chern-Simons term is only gauge-invariant up to a boundary term.  So any theory based purely on the axion coupling for an internal gauge field cannot generate a dynamical boundary from the requirement of gauge invariance in the way that the Chern-Simons theory does.

Our approach leads to a version of $BF$ topological field theory in two and three dimensions as the effective description of topological insulators.  This theory, which we introduce starting with the two-dimensional (2D) case in the following section, preserves time-reversal and parity symmetries, unlike the Chern-Simons theory; in particular  It also can exist in three dimensions (3D), again unlike the Chern-Simons theory.  In two dimensions, it is well known that the relevant $BF$ theory can be recast as two copies of Chern-Simons theory~\cite{HanssonBF,ReviewBF}, but we will argue that the $BF$ picture captures better the important properties of the 2D topological insulator, in particular its intrinsic time-reversal invariance.

We obtain the 2D $BF$ theory heuristically in the following section starting from the distinctive response of a 2D topological insulator to an electromagnetic $\pi$ flux (i.e., insertion of a solenoid with flux $h c / 2 e$)~\cite{fu&kane2-2006,essinmoore,ranvishwanathlee,qizhangpi}.  In 3D, the $BF$ theory involves one vector gauge field $a_\mu$ and one rank-two tensor field $b_{\mu \nu}$.  These fields are associated with the distinctive response of a 3D topological insulator to a flux {\it line}~\cite{franzwormhole}, which is closely related to the response of the ``weak'' topological insulator to a dislocation~\cite{ranzhangvishwanath}.  So in both 2D and 3D we can understand $BF$ theory from the attachment of nontrivial responses to electromagnetic flux.

Section III introduces some general features of $BF$ theories of relevance to topological insulators, and Section IV provides more details for the 2D case.  Section V contains most of the main results of this paper on the 3D case.  One type of electromagnetic coupling is present even with time-reversal symmetry, while breaking time-reversal symmetry at the boundary is shown to lead to the ${\bf E} \dot {\bf B}$ term after integrating out the internal fields of $BF$ theory.  The electromagnetic current contains both ``charge'' and ``polarization'' contributions, and we obtain the ``wormhole effect''~\cite{franzwormhole} along flux lines with gapped surfaces as a consequence of the coupling.  The existence of the Dirac fermion at surfaces when time-reversal-symmetry is present, which is the last result in Section V, becomes more understandable on noting that the combination of scalar and vector bosons that the bulk $BF$ theory implies at the surface are precisely those required to represent a Dirac fermion via the explicit ``tomographic'' mapping in 2+1D~\cite{Bosonization1,Bosonization3}.

Some generalizations and future directions are reviewed in Section VI; one is that the $BF$ theory description allows for fractional statistics not just in 2D but in 3D~\cite{HanssonBF,ReviewBF,StatisticsBF,StatisticsDegeneracyBF}.  This does not contradict the classic argument that pointlike objects do not have fractional statistics in 3D~\cite{leinaasmyrheim} because the fractional statistics are between one pointlike and one stringlike object. As yet there is no microscopic realization of a ``fractional'' topological insulator (we discuss some issues in closing), but the $BF$ approach can at least formally describe such states.  Rather than try to give an overview of the other features of $BF$ theory at the outset, it seems desirable to proceed with deriving it from the $\pi$ flux response in the simplest case, then explain how other features of topological insulators arise in the $BF$ theory.

\section{Vortices with $\pi$ flux in a 2D topological insulator}


In topological insulators, an electromagnetic $\pi$ flux has special roles both in 2D and 3D. When $\pi$ flux is threaded, there are always some excitations which are stable to any local $T$- symmetric perturbation. In 2D TIs, $\pi$ flux can induce spin-charge separation with semionic statistics. In 3D TIs, $\pi$ flux creates helical 1D metals. Since this property is topological, $\pi$ flux response can be used to distinguish TIs from trivial phases without referring to edge states. However, $\pi$ flux is a strong perturbation from the point of view of an electron because the electron wavefunction gains a  $(-1)$ phase factor when it is rotated around the flux.  If our goal is to depict the TI as a condensed phase, as in the early work on the Chern-Simons description of the quantum Hall effect, then for the condensate to consist of charge $e$ objects, the phase factor must be somehow compensated if the $\pi$ flux is to be a finite-energy excitation.


So there must be another global effect that compensates the $(-1)$ phase factor for electrons when $\pi$ flux is threaded.  We will show that this is the origin of a BF-type action.
 
\subsubsection{Review of the $2\pi$ flux vortex}
Let's consider a vortex threaded into a 2D condensate of charge $e=1$ objects (electrons).  Consider the condensate wave function $\psi$ and a $U(1)$ electromagnetic gauge field $A$ in a vortex configuration. The Lagrangian density for the single vortex follows $(\hbar = c = 1)$:
\begin{equation}
L = \frac{1}{2}\mid(\partial_{\mu}-iA_{\mu})\psi\mid^{2} - V(\psi^{\dagger}\psi),\qquad V = \frac{1}{2}(\psi^{\dagger}\psi -v^{2})^{2} 
\label{eq1}
\end{equation} 
The minimum energy configuration for $\psi$ is easy to obtain : $\psi = v \exp(i\theta)$ where $\theta$ is the angle with respect to $x$-axis in 2D. This distribution of $\psi$ has two key features. First, it minimizes the vortex potential $V$. Second, $\psi$ is single-valued because of the angle dependence~ $\exp(i\theta)$. And if we substitute this distribution of $\psi$ into equation (\ref{eq1}), we obtain just
\begin{equation}
L = \frac{v^{2}}{2}(\partial_{\mu}\theta - A_{\mu})^{2}
 \label{eq2}
 \end{equation}
The equation (\ref{eq2}) fixes the field configuration of $A$ to make the system have finite energy: $A_{i}$ should be reduced into $\partial_{i}\theta$ as $r \rightarrow \infty$. Otherwise, the energy of the vortex is simply infinite which is not physically acceptable. Due to this constraint, the flux trapped in the vortex configuration is quantized by $2\pi$:
\begin{equation}
\int{d^{2}x \varepsilon_{ij}\partial_{i}A_{j}} = \oint d\overrightarrow{x} \centerdot \overrightarrow{A} = \triangle \theta = 2\pi
\end{equation} 
In general, the single-valuedness requires $\psi \propto \exp(in\theta)$ which corresponds to an $n$-vortex configuration. Then this gives the total flux $2n\pi$ threaded in the 2D condensate. Hence, we directly see that $\pi$ flux is not allowed for the vortex configuration with finite energy. Thus, we need another component in the theory to deal with $\pi$ flux vortex. 

\subsubsection{$\pi$ flux vortex and $BF$ theory}
\begin{figure}
\includegraphics[width=0.75\columnwidth]{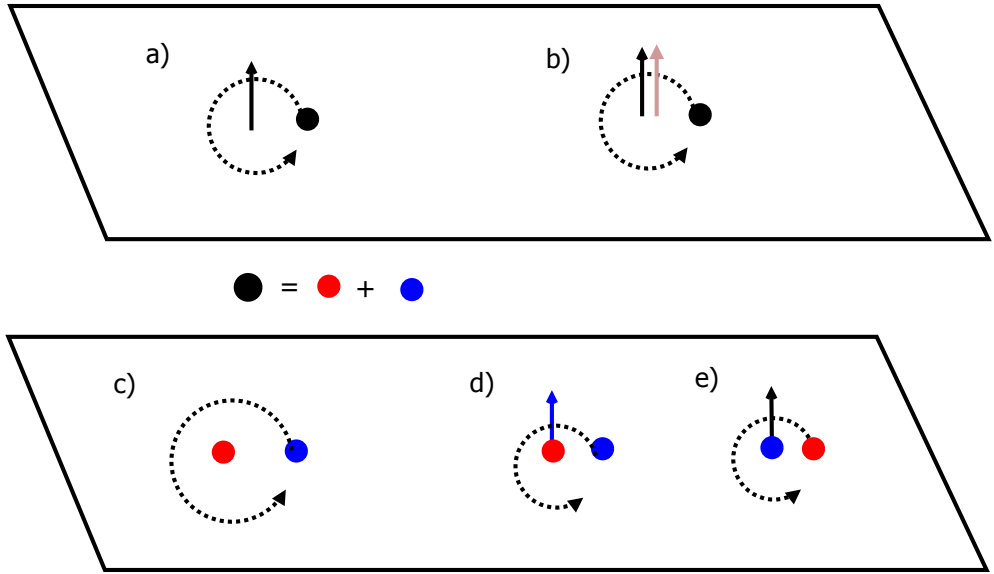}
\caption{The black arrows represent one-half magnetic flux quantum (a $\pi$ flux) and the black spheres represent electrons.  Consider a system with $S_z$ conservation for simplicity: an electron can be thought as a composite of a holon (red spheres) and an $S_z = \pm \frac{1}{2}$ spinon (blue spheres). a) Whenever an electron is circulated around the half flux quantum ($\pi$ flux), the electron acquires a phase factor $(-1)$ which shifts the energy levels of electrons. b)This phase factor can be cured by introducing a statistical gauge field $a$ which is tied to the EM flux tubes; the $BF$ term does this job.  Hence the electrons see an effective $2\pi$ flux and behave as though there is no flux tube present. c) Spinons and holons have relative semionic statistics and thus every holon is affected in the same way as an electron whenever there is a spinon present. This effect generates the phenomenon that EM $\pi$ flux captures one spinon and $\pi$ flux of the fictitious spin-gauge field captures one holon, as illustrated in d) and e).}
\label{Fig1}
\end{figure}

As explained in the previous section, a finite-energy vortex configuration with $\pi$ flux is seemingly excluded for a charge-$e$ condensate. This result can be traced back to the phase factor $(-1)$ for electrons rotating around $\pi$ flux. Thus, if there is a way to create another phase factor $(-1)$ for electrons, then $(-1)(-1) = 1$ would allow $\pi$ flux in the vortex configuration with finite energy. This is done by introducing a $U(1)$ statistical gauge field $a_\mu$ that couples to electrons. This is depicted in the figure (\ref{Fig1}). Now the Lagrangian (\ref{eq1}) is replaced by the following equation.
\begin{equation}
L = \frac{1}{2}\mid(\partial_{\mu}-i(A_{\mu}+a_{\mu}))\psi\mid^{2} - V(\psi^{\dagger}\psi)\qquad V = \frac{1}{2}(\psi^{\dagger}\psi -v^{2})^{2}
\label{eq3}
\end{equation} 
Both gauge fields $A$ and $a$ carry flux $\pi$ for the simplest case. However, $a$ can carry $(2n-1)\pi$ flux (or equivalently, $\pi$ flux of $a$ induces phase factor $\exp(i(2n-1)\pi)=(-1)$) because the single-valuedness of the wave function only requires the total phase change to be $2n\pi$ in rotating around vortex. For the purpose of deriving the BF-type action, $n=1$ is enough, and we will stick to $n=1$ in this section, but the generalization to $n\neq 1$ is trivial, and the importance for $n\neq 1$ will be pointed out later when we discuss the statistics of$BF$theory and the $\mathbb{Z}_{2}$-ness of TIs. With the same ansatz $\psi = v \exp(i\theta)$ as before, we get the Lagrangian density for the gauge fields $a$ and $A$:
\begin{equation}
L = \frac{v^{2}}{2}(\partial_{\mu}\theta - A_{\mu}-a_{\mu})^{2}
\label{eq4}\end{equation} 
We can use the standard dual representation for the vortex by introducing an auxiliary $U(1)$ field $\xi^{\mu}$. Then the Lagrangian density (\ref{eq4}) can be written as the following:
\begin{equation}
L = -\frac{1}{2v^{2}} {\xi^{\mu}}^{2} + \xi^{\mu}(\partial_{\mu}\theta- A_{\mu}-a_{\mu})
\end{equation}
Now the key step of the dual representation follows: writing $\theta = \theta_{smooth} + \theta_{vor}$. The integration (or gauging away) of $\theta_{smooth}$ gives the constraint $\partial_{\mu} \xi^{\mu} = 0$. This can be solved by introducing another $U(1)$ gauge field $b$ such that $\xi^{\mu} = \varepsilon^{\mu\nu\lambda}\partial_{\nu}b_{\lambda}$, and this transformation is enough to show there should be a BF-type term in the Lagrangian to deal with $\pi$ flux vortex. After using the relation between $\xi$ and $b$, we obtain 
\begin{equation}
L = -\frac{1}{4v^{2}} f_{\mu\nu}^{2} + \varepsilon^{\mu\nu\lambda} \partial_{\nu} b_{\lambda}(\partial_{\mu}\theta_{vor}- A_{\mu}-a_{\mu}),
\end{equation}
where $f$ is the field strength of $b$. Finally, we identify the vortex current as $j^{\mu}= \frac{1}{2\pi} \varepsilon^{\mu\nu\lambda}\partial_{\nu}\partial_{\lambda}\theta_{vor}$, and write the Lagrangian in a slightly different form:
\begin{equation}
L = -\frac{1}{4v^{2}} f_{\mu\nu}^{2} + 2\pi b_{\mu} j^{\mu} - \varepsilon^{\mu\nu\lambda}(A_{\mu}+a_{\mu})\partial_{\nu}b_{\lambda}
\label{eq5}
\end{equation}
Now we note that the first two terms in the equation (\ref{eq5}) are induced by the vortex configuration, and argue that the effective theory for the background TI should be the third term including two gauge fields $a$ and $b$ with non-dynamical external gauge field $A$.  Another way to motivate discarding the first term is that, when we deal with long-range and low-energy physics, we can ignore this Maxwell-like term for the same reason that pure Chern-Simons theory is the effective theory for quantum hall systems: the third term in equation \ref{eq5} includes only one derivative, whereas the Maxwell-like term carries two derivatives and is relatively small at long length scales. With slightly different normalization, we obtain the effective theory for the 2D topological insulator: 
\begin{equation}
L_{BF} = \frac{1}{\pi} \varepsilon^{\mu\nu\lambda}(A_{\mu}\partial_{\nu}b_{\lambda}+a_{\mu}\partial_{\nu}b_{\lambda})
\label{eq6}
\end{equation}
The first term in equation (\ref{eq6}) is current-gauge coupling where the current is mediated by the gauge field $b$.   The second term is called the $BF$ term (because it includes the gauge field $b$ and the field strength of the $a$ field), and it contains the coupling between $a$ and $b$. In fact, the $BF$ term is topological, i.e it does not carry dynamics; rather it carries the information of statistics between sources of $a$ and $b$.  Note that the first term is actually a $BF$ term as well, coupling the internal field $b$ and the external field $A$.  This type of mixed $BF$ term has previously been obtained for the 2D topological insulator by Goryo et al.~\cite{goryo1,goryo2,goryo3}.  The meaning of the coefficients and the $BF$ term will be clarified in the next section, and in this section we focus on consequences of (\ref{eq6}).  There is an interesting property for the equation (\ref{eq6}): if we integrate out $a$ and $b$, then the resulting action is $0$, i.e., the bulk effective action for $A$ is $0$. So if we want to have an electromagnetic response due to the external gauge field $A$, we need another term coupled to $a$, as in the $T$-breaking Chern-Simons term for the quantum Hall effect.  Another difference is that the $BF$ theory with $A$ above preserves $P$- and $T$- symmetry which will be clear in the next section. Looking ahead, a key feature of this $BF$ theory is that it has a direct 3D generalization; for 3D, we will similarly derive the existence of a two-form gauge field $b$ coupled to $a$ and $A$,
\begin{equation}
L_{BF} = \frac{1}{2\pi} \varepsilon^{\mu\nu\lambda\rho}(A_{\mu}\partial_{\nu}b_{\lambda\rho}+a_{\mu}\partial_{\nu}b_{\lambda\rho})
\label{eq7}\end{equation}
Then the equation (\ref{eq7}) inherits the properties of 2D $BF$ theory (\ref{eq6}): it is topological, the effective action of $A$ is zero, and it is $P$- and $T$- symmetric.  

\section{General properties of $BF$ theory}

Our goal in this paper is to argue that $BF$ theory is the effective theory for TIs in a way similar to CS theory for the quantum Hall effect.  Before continuing with that argument, let us review a few more properties of $BF$ theory. $BF$ theory is a generalization of CS theory, and both are topological~\cite{HanssonBF,ReviewBF,DegeneracyBF2}; in fact, the relationship between $BF$ and CS theory is fairly simple in (2+1)D.  This relationship is essentially the same as the microscopic relationship between (two copies of) the integer quantum Hall effect and the 2D TI.  But there are important differences between $BF$ and CS theory, such as symmetry; some similarities and differences between CS and $BF$ theory in two and three spatial dimensions will be reviewed in the following.  Versions of $BF$ theory have been applied in condensed matter physics previously to study superconductors and topological spin liquids~\cite{HanssonBF,DegeneracyBF2,YeBF}, but not topological insulators as far as we are aware.  Henceforth we follow the condensed matter convention and refer to systems by their spatial dimension, e.g., 2D means (2+1)D.

\subsection{2D $BF$ theory and CS theory}
The most general form of $BF$ theory in 2D is composed of two $U(1)$ gauge fields $a$, $b$ and corresponding sources $j$, $J$~\cite{HanssonBF,ReviewBF}. The full $BF$ action is 
\begin{equation}
L_{BF} = \frac{k}{2\pi} \varepsilon^{\mu\nu\lambda}a_{\mu}\partial_{\nu}b_{\lambda} - j^{\mu}a_{\mu} - J^{\mu}b_{\mu}
\label{7}  
\end{equation}
The first term in equation \eqref{7} is topological and describes braiding statistics with parameter $k$ : when the quasiparticle type $j$ is rotated around the quasiparticle type $J$, the total wave function obtains $\frac{2\pi}{k}$. This term has the same role as the term $L_{CS} = \frac{k}{4\pi}\varepsilon^{\mu\nu\lambda}a_{\mu}\partial_{\nu}a_{\lambda}$ in CS theory, and the equations of motion are also simply obtained:
\begin{equation}
\frac{k}{2\pi} \varepsilon^{\mu\nu\lambda}\partial_{\nu}b_{\lambda} = j^{\mu} \qquad 
\frac{k}{2\pi} \varepsilon^{\mu\nu\lambda}\partial_{\nu}a_{\lambda} = J^{\mu}
\end{equation}
The symmetries of the gauge fields $a$ and $b$ are different: under $T$, $(a_{0}, a_{1}, a_{2})\rightarrow (a_{0}, -a_{1}, -a_{2})$ and $(b_{0}, b_{1}, b_{2})\rightarrow (-b_{0}, b_{1}, b_{2})$. So the $BF$ term in the equation (\ref{eq7}) is $T$ - invariant. Armed with these, we reconsider the effective theory (\ref{eq6}):
\begin{equation}
L_{BF} = \frac{1}{\pi} \varepsilon^{\mu\nu\lambda}(A_{\mu}\partial_{\nu}b_{\lambda}+a_{\mu}\partial_{\nu}b_{\lambda})
\end{equation}
Two comments about this theory are in order. First, because $(A_{0}, A_{1}, A_{2})\rightarrow (A_{0}, -A_{1}, -A_{2})$ under $T$- transformation (the same symmetries as the $a$ gauge field, which is why it is natural to describe the $\pi$ vortex by combining $A$ and $a$), we see that the theory is $T$- invariant, i.e., the minimal coupling between $A$ and $b$ is $T$- symmetric.  $A$ acts as a source of $b$ field. Secondly, our theory for topological insulators is described by $k=2$, and this makes sense because $\pi$ flux of $a$ field should give the phase factor $(-1)$ to the total wave function as we discussed $\pi$ flux vortex theory(In fact, $k=2$ corresponds to the semionic statistics for $a$ and $b$). However, this is not the only coefficient giving phase factor $(-1)$ between $a$ and $b$. In general, $k=\frac{2}{2n-1}$ should be equally good because $\exp(i(2n-1)\pi)=(-1)$. So the most general theory with $\pi$ flux having the proper statistical effect has the form : 
\begin{equation}
L_{BF} = \frac{1}{(2n-1)\pi} \varepsilon^{\mu\nu\lambda}a_{\mu}\partial_{\nu}b_{\lambda} + \frac{1}{\pi} \varepsilon^{\mu\nu\lambda}A_{\mu}\partial_{\nu}b_{\lambda}
\end{equation}

For any value of $n$, the effective action for $A$ is still $0$, so there is no (bulk) response to the external field $A$.  
Before getting into the details of the system's response, it should be noted that the action of 2D $BF$ theory is just doubled CS theory with opposite chirality; while it is known that the 2D TI can be realized as the double copy of quantum hall layers with opposite chirality, the above theory is saying that this is true for some properties, such as the existence of edge modes, even when there is no conserved component of spin. The equation \eqref{eq6} can be equally well derived if we start from doubled CS theory and pick appropriate combinations of the two CS fields (``charge'' and ``spin''). However, it is well known that TIs can be described only by doubled CS theories describing odd-integer quantum Hall effects, and we wish to emphasize here that this is really coming from the $\pi$ flux response, which is a defining property of TIs, and not from the assumption of decoupled spin-up and spin-down layers.

\subsubsection{3D $BF$ theory}  
The most general form of 3D $BF$ theory includes a one-form gauge field $a_{\mu}$, an antisymmetric two-form gauge field $b_{\mu\nu}$, and corresponding sources  $j^{\mu}$, $\Sigma^{\mu\nu}$~\cite{HanssonBF,StatisticsBF,ReviewBF,StatisticsDegeneracyBF}. Here, the source $\Sigma$ is an antisymmetric rank 2 tensor, so physically the source $\Sigma$ represents the density of line-like objects that can be pictured as field lines for electric and magnetic parts of the tensor.
We will later connect the magnetic and electric components of $\Sigma$ through the equations of motion.   The corresponding $BF$ theory should contain information about braiding of vortices and quasiparticles. Interestingly, 3D $BF$ theory can be used to describe fractional statistics in 3D systems.  The standard argument that pointlike particles cannot have fractional statistics in 3D as they do in 2D~\cite{leinaasmyrheim} is that all ways of looping one particle around another to detect their mutual statistics are topologically equivalent in 3D but not in 2D; as a result, 3D statistics are described by the permutation group, and 2D statistics are described by the much richer braid group--one consequence is that in 2D, there are ``anyonic'' particles that are neither fermionic nor bosonic.  In 3D, vortices and quasiparticles can braid noncontractibly like particles in 2D, allowing for statistics other than fermionic and bosonic.




Hence it is possible to impose nontrivial statistics between vortices and quasiparticles, and we will argue that this is what happens in 3D TIs. The starting $BF$ theory is the following: 
\begin{equation}
L_{BF} = \frac{k}{4\pi} \varepsilon^{\mu\nu\lambda\rho}a_{\mu}\partial_{\nu}b_{\lambda\rho} - j^{\mu}a_{\mu} - \frac{1}{2}\Sigma^{\mu\nu}b_{\mu\nu}
\label{8}
\end{equation}
As in the 2D case, the first term carries the statistics: when quasiparticle $j$ is rotated around the vortex density $\Sigma$, the total wave function obtains a phase factor $\frac{2\pi}{k}$.  The equation of motion can be obtained by varying with respect to $a$ and $b$ : 
\begin{equation}
\frac{k}{4\pi} \varepsilon^{\mu\nu\lambda\rho}\partial_{\nu}b_{\lambda\rho} = j^{\mu} \qquad
\frac{k}{2\pi} \varepsilon^{\mu\nu\lambda\rho}\partial_{\lambda}a_{\rho} = \Sigma^{\mu\nu}
\end{equation}
And as in 2D, $k=2$ (or more generally, $k=\frac{2}{2n-1}$) will describe a TI for which threading with $\pi$ flux yields the correct phase factor $(-1)$. The external field $A$ acts as a source for the two-form field $b$ if we look into the equation \ref{7}. Finally, this $BF$ theory is $T$- symmetric because $(a_{0}, a_{1}, a_{2},a_{3})\rightarrow (a_{0}, -a_{1}, -a_{2}, -a_{3})$, $(b_{0i}, b_{ij})\rightarrow (-b_{0i}, b_{ij})$ and $(A_{0}, A_{1}, A_{2}, A_{3})$ $\rightarrow$ $(A_{0}, -A_{1}, -A_{2}, -A_{3})$. Thus the minimal coupling between $A$ and $b$ is still $T$- symmetric, and the candidate theory for 3D TIs with integer $n$ is following : 
\begin{equation}
L_{BF} = \frac{1}{2(2n-1)\pi} \varepsilon^{\mu\nu\lambda\rho}a_{\mu}\partial_{\nu}b_{\lambda\rho} +\frac{1}{2\pi} \varepsilon^{\mu\nu\lambda\rho}A_{\mu}\partial_{\nu}b_{\lambda\rho} 
\label{9}
\end{equation}
As noted in 2D $BF$ theory, we need a term coupled to $a$ to have a total bulk response of external gauge field $A$, which will be obtained later.

\section{2D topological Insulators}
The physics of 2D TIs is relatively easy to understand compared to that of 3D TIs because a 2D TI can be realized as a bilayer integer quantum hall system with opposite $T$- symmetry, although this picture implicitly assumes a $U(1)$ spin rotation symmetry and fails to capture the $\mathbb{Z}_2$ nature of the topological invariant.  The role of the magnetic field in ordinary quantum hall systems is replaced by the spin-orbit interaction, whose symmetries require opposite integer quantum Hall states of up and down spins along the $U(1)$ axis.
When there is no $U(1)$ remnant of spin-rotation symmetry, we are forced to interpret $B$ as so-called time-reversal gauge field, and $J$ as the time-reversal-pair current. In fact, the above Lagrangian for 2D TIs is very similar to that of quantum hall systems, and many aspects of physics in 2D TIs resemble those from quantum Hall states.  We quickly review the topological field theory description of quantum hall systems. 

\subsection{Quantum Hall systems and CS theory}         
Let us consider a $\nu=\frac{1}{3}$ fractional quantum hall system, i.e., the lowest Landau level is 1/3 full. The system has elementary excitations with fractional charges : fractionalized quasiholes($=\frac{e}{3}$) and quasielectrons($=-\frac{e}{3}$). These excitations can be made in the system if we thread magnetic flux, and the braiding statistics of those quasiparticles are fractional. Remarkably, these seemingly complicated properties can be simply formulated in CS theory :   
\begin{equation}
L_{CS} = -\frac{k}{4\pi}\varepsilon^{\mu\nu\lambda}a_{\mu}\partial_{\nu}a_{\lambda} + j^{\mu}a_{\mu}, \qquad j^{\mu} = \frac{1}{2\pi}\varepsilon^{\mu\nu\lambda}\partial_{\nu}A_{\lambda}
\end{equation} 
Here $k$ parametrizes the braiding of quasiparticles and filling factor $\nu=\frac{1}{3}$ is described $k=3$. The requirement of gauge invariance for the total action on a manifold with boundary, with $L_{CS}$ as bulk action, leads to an edge chiral boson field which is indeed present in quantum hall systems~\cite{wenrev1,Wen95}. Moreover, if we integrate out the internal gauge field $a$, we obtain the correct effective electromagnetic Lagrangian for a state with Hall effect $\sigma_{xy} = e^2/3 h$: 
\begin{equation}
L_{QHE} = -\frac{1}{12\pi}\varepsilon^{\mu\nu\lambda}A_{\mu}\partial_{\nu}A_{\lambda}
\end{equation}
Thus we can argue that CS theory is a valid effective theory for quantum hall systems due to its correct description of statistics of quasiparticles, correct effective theory for $A$, and edge theory. In the next and subsequent sections, we will argue $BF$ theory also captures statistics of quasiparticles, correct effective theory for $A$ and $B$, and edge theory of 2D TIs, i.e., $BF$ theory should be the correct effective theory for TIs. 

\subsection{2D topological insulators and $BF$ theory}
Topological insulators are gapped topologically ordered systems like quantum Hall states, and thus it is natural that they could be described by a topological field theory. However, $BF$ theory, unlike the simplest form of CS theory, involves two entities in the theory $a$ and $b$ with different symmetries. If we want to follow the philosophy of CS theory, then $a$ and $b$ should couple to currents of elementary excitations in the system.  So it would be good to think of possible excitations in TIs before moving into detailed mathematics.

To construct an analogy between CS theory and $BF$ theory, we will stick to 2D TIs in this section. In a quantum hall system, we can collect the elementary excitation by threading the magnetic flux. Similarly, we can thread a flux of the external electromagnetic gauge field into a 2D TI, and it is known that we can have two types of excitations in the presence of a $\pi$ flux~\cite{ranvishwanathlee,qizhangpi}. In 2D TIs, the excitations are a charge-neutral Kramers doublet (charge $0$ and, if one component of spin is a good quantum number, spin$=\pm \frac{1}{2}$) and holon(charge$=\pm e$ and spin zero).  So $a$ and $b$ should couple to the sources of external gauge fields which are currents of ``spinons'' and ``holons''; of course the external gauge field coupling to the Kramers-doublet is not physical, and the other gauge field is ordinary $U(1)$ electromagnetism.

Note that even if no $U(1)$ component of spin rotation symmetry like $S_z$ is conserved microscopically, there are still two $U(1)$ currents macroscopically, not just the $U(1)$ charge current.  This is reflected in the existence of two dissipationless edge modes that are asymptotically decoupled at low energy, and the generic existence of a dissipationless ``spin Hall effect'' carried by edge states in an applied electric field even if the direction and magnitude of this spin Hall effect are not quantized.  In the $BF$ description, there are two $U(1)$ currents of which one couples directly to the external electromagnetic field; the quantized particle source of the ``spin'' gauge field can be visualized as one of the charge-neutral excitations created by insertion of a $\pi$ flux.

We recall the previous $BF$ theory written for 2D TIs : 
\begin{equation}
L_{BF} = \frac{1}{\pi} \varepsilon^{\mu\nu\lambda}(A_{\mu}\partial_{\nu}b_{\lambda}+a_{\mu}\partial_{\nu}b_{\lambda})
\end{equation} 
From the above equation, we see that $b$ couples to $A$ and thus represents the field for holons. Naturally, $a$ should represent spinons. More precisely, the current for $a$ couples to a fictitious time-reversal gauge field $B$; the term for the coupling between spinons and external time-reversal gauge fields was absent in the the previous discussions because we only dealt with the physical EM field.  To make the duality between $a$ and $b$ in 2D explicit, we now add $\frac{1}{\pi} \varepsilon^{\mu\nu\lambda} B_{\mu}\partial_{\nu}a_{\lambda}$ to our Lagrangian density. Then the full Lagrangian for our system is obtained : 
\begin{equation}
L_{BF} = \frac{1}{\pi} \varepsilon^{\mu\nu\lambda}a_{\mu}\partial_{\nu}b_{\lambda} +\frac{1}{\pi} \varepsilon^{\mu\nu\lambda}A_{\mu}\partial_{\nu}b_{\lambda}+\frac{1}{\pi} \varepsilon^{\mu\nu\lambda}B_{\mu}\partial_{\nu}a_{\lambda}
\label{10}
\end{equation}    
The equations of motion for $a$ and $b$ can be obtained if we vary $b$ and $a$. 
\begin{equation}
\varepsilon^{\mu\nu\lambda}\partial_{\nu}b_{\lambda} = -\varepsilon^{\mu\nu\lambda}\partial_{\nu}B_{\lambda}, \qquad
\varepsilon^{\mu\nu\lambda}\partial_{\nu}a_{\lambda} = -\varepsilon^{\mu\nu\lambda}\partial_{\nu}A_{\lambda}
\end{equation}
From the above equations, some important features can be directly read off. First, unit spinon and holon are localized at the $\pi$ flux of $A$ and $B$ from the equation of motion. Second, this spinon and holon obey semionic statistics. Thus, the above $BF$ theory \eqref{10} correctly describes the elementary excitations in 2D TIs. Now, consider the effective Lagrangian for the external fields $A$ and $B$ obtained by integrating out $a$ and $b$. By completing squares for $a$ and $b$ and integrating out them, we obtain the following effective Lagrangian: 
\begin{equation}
L_{eff} = \frac{1}{\pi} \varepsilon^{\mu\nu\lambda} A_{\mu}\partial_{\nu}B_{\lambda}
\end{equation}
Note that there is no bulk electromagnetic action for $A$ alone, consistent with the absence of a charge Hall effect; the effective action for $A$ and $B$ in bulk describes a ``quantum spin Hall effect''  (a spin current flows in response to a charge field and vice versa) if there is a conserved current of spin in some direction.  Since we can trivially rewrite the bulk $BF$ theory as a doubling of CS theory in 2D, it contains two chiral bosons at the edge, consistent with 2D TIs.  We will review the derivation of this edge theory later in Section V as preparation for the surface theory of the 3D TI.


Before finishing this section, it is worth mentioning briefly how the $BF$ theory can capture the physics of the ``$Z_2$ odd'' property of 2D TIs.  Consider an $N$-flavored $BF$ theory to describe $N$ copies of 2D TI 
\begin{equation}
L= \sum_{i=1}^{i=N}(\frac{1}{\pi} \varepsilon^{\mu\nu\lambda} a^{i}_{\mu}\partial_{\nu}b^{i}_{\lambda}+\frac{1}{\pi} \varepsilon^{\mu\nu\lambda} A_{\mu}\partial_{\nu}b^{i}_{\lambda} + \frac{1}{\pi} \varepsilon^{\mu\nu\lambda} a^{i}_{\mu}\partial_{\nu}B_{\lambda})
\end{equation} 
This edge of this Lagrangian can be studied from the $K$-matrix of the system. The $K$-matrix for the single-flavor $BF$ theory is $K$ = diag$(1,-1)$ when the $BF$ term is diagonalized. This $K$-matrix captures that there are two one-dimensional edge modes with opposite chiralities. When there are $N$ flavors, the $K$-matrix is trivially generalized as $K$ = diag$(1,-1,1,-1,1,-1\cdots)$ where the block diag$(1,-1)$ is repeated $N$ times. This edge is unstable to $T$-symmetric perturbations if $N$ is even. Thus, we are led to conclude $N=(2n-1)$. This $(2n-1)$-flavor 2D TI can be fit into a single-component $BF$ theory if we consider only the response to the external gauge fields. (That is, if we only look at the overall response to the external gauge fields, we do not actually need to retain the index $i$ which labels different flavors, but can consolidate all flavors into one.) Then we are led to the simplified Lagrangian 
\begin{equation}
L= (2n-1)\times(\frac{1}{\pi} \varepsilon^{\mu\nu\lambda} a_{\mu}\partial_{\nu}b_{\lambda}+\frac{1}{\pi} \varepsilon^{\mu\nu\lambda} A_{\mu}\partial_{\nu}b_{\lambda} + \frac{1}{\pi} \varepsilon^{\mu\nu\lambda} a_{\mu}\partial_{\nu}B_{\lambda})
\label{nSHE}
\end{equation} 
This theory has a $(2n-1)$-times amplified spin Hall effect. This theory still has a well-defined response to the $\pi$ fluxes as $\pi$ flux is still the smallest $T$-symmetric flux. Hence, we still want to treat $\pi$ fluxes as the unit sources of $a$ and $b$. This means we need to rescale the gauge fields $a$ and $b$: $(2n-1)a \rightarrow a$ and $(2n-1)b \rightarrow b$. This scaling applied to the Lagrangian (\ref{nSHE}) gives 
\begin{equation}
L= \frac{1}{(2n-1)\pi} \varepsilon^{\mu\nu\lambda} a_{\mu}\partial_{\nu}b_{\lambda}+\frac{1}{\pi} \varepsilon^{\mu\nu\lambda} A_{\mu}\partial_{\nu}b_{\lambda} + \frac{1}{\pi} \varepsilon^{\mu\nu\lambda} a_{\mu}\partial_{\nu}B_{\lambda}
\end{equation} 
which has been seen before when we considered $\pi$-flux responses. 

In conclusion, $BF$ theory can describe key aspects of 2D TIs: the importance of $T$- symmetry, the elementary excitations and their statistics, and the edge theory.

\section{3D topological insulators}

3D topological insulators are described by four $\mathbb{Z}_2$ topological invariants, one ``strong'' and three ``weak''.   Topological insulators with only nontrivial weak invariants can be adiabatically connected to layering of 2D topological insulators and are unstable to disorder. On the other hand, the strong topological insulator is an intrinsically 3D system and does not have a two-dimensional limit.  Moreover, the essence of the 2D topological insulator can be understood relatively easily by considering ``quantum spin Hall'' systems where a $U(1)$ subgroup of spin rotation symmetry is preserved.  However, if we try to preserve a $U(1)$ spin rotation in 3D topological insulators, we end up with only weak topological insulators rather than the strong topological insulator. This makes it difficult to understand the strong topological insulator because we need information about spins in the system but we cannot identify states by a remaining spin quantum number.

BF theory supplies a route around this difficulty.  Roughly speaking, the $b$ and $a$ fields in $BF$ theory carry information about spin and charge and couple to the external electromagnetic gauge field $A$. In this section, we will figure out what $a$ and $b$ should be on physical grounds and justify that our version of $BF$ theory is a correct effective description of 3D strong topological insulators: it captures the quantized magnetoelectric effect~\cite{qilong,essinmoorevanderbilt} and the braiding of excitations in the system when the surfaces are gapped, and as mentioned in the Introduction it also allows the possibility of gapless fermionic surfaces.

\subsection{3D topological insulators with gapped surfaces and $BF$ theory}

As we did above for the 2D TI and $BF$ theory, we need to identify the coupling for the internal gauge field $a$. The $a$ field is still a one-form even in (3+1)D $BF$ theory, and the external gauge field $A$ is also a one-form. So the simplest coupling would be $\sim da \wedge dA$, but this breaks $T$- symmetry of the system as $T$ takes $da \wedge dA \rightarrow -(da \wedge dA) $. However, note that the term is a total derivative and hence can be sourced from a surface where its coefficient changes, i.e $da \wedge dA = d(a\wedge dA)$. Thus, we are led to conclude that our theory describes TI with broken $T$- invariance only on the surface.  As before, we also have $T$- invariant couplings between $b$ and $a$, $A$. The resulting theory contains three terms.  
\begin{equation}
L_{BF} = \frac{1}{2\pi} \varepsilon^{\mu\nu\lambda\rho}a_{\mu}\partial_{\nu}b_{\lambda\rho} +\frac{1}{2\pi} \varepsilon^{\mu\nu\lambda\rho}A_{\mu}\partial_{\nu}b_{\lambda\rho} +C\varepsilon^{\mu\nu\lambda\rho} \partial_{\mu}a_{\nu}\partial_{\lambda}A_{\rho}
\label{12}  
\end{equation}

We will find $Z_{2}$-ness, the existence of only two phases as a result of $T$ invariance, by studying the third term in the theory. To proceed further, we need an extra piece of information which is not encoded in the field theory itself: the gauge charge lattice. This is because the response to EM field is sensitive to the quantization of charges in the system. For convenience, we deal with a compact (3+1)D manifold such as $T^{4}$. On such manifolds, we have no boundary and hence the third term might be thought to be trivial. However, this is not the case as we will consider $T$- invariance of the theory. We separate the terms in the theory \eqref{12} as follows : $L_{\rm bulk} =\frac{1}{2\pi} \varepsilon a\partial b +\frac{1}{2\pi} \varepsilon A\partial b$ and $L_{surf} = C \varepsilon \partial a \partial A $ where anti-symmetrization is done implicitly. The full theory is $\exp(i\int (L_{\rm bulk} + L_{\rm surf}))$. Under the time-reversal symmetry operation, we obtain $\exp( i\int (L_{\rm bulk}-L_{\rm surf}))$. Hence, for the theory to be well-defined and have $T$- symmetry, we need to have $\exp(i\int L_{\rm surf})=\pm 1$. 

Now, we impose the gauge charge lattice for the theory. First, there is a well-defined particle excitation $j_{a}^{\mu}$ which is the source for the gauge $a$. We quantize $j_{a}^{0}$ to be integral over each of the 2D tori within $T^4$. Then from the second term in $L_{\rm bulk}$, we are led to conclude that $J_{em}^{0}$ is also quantized to be an integral multiple of $e$.  
This seemingly trivial imposition of the gauge charge lattice gives a nontrivial effect on $L_{\rm surf}$ as the fluxes threaded in each 'torus' of the manifold should be quantized by $\frac{2\pi}{e}=2\pi$ (here $e=1$ in our units). Then $L_{\rm surf}$ can take only the following possible values. 
\begin{equation}
\int L_{\rm surf} = C 2\pi Z \times 2\pi Z \times 2 = C 8\pi^{2} Z.
\end{equation}
Here $Z$ indicates an unknown integer. The first $2\pi Z$ is due to the integral of $\partial A$ and the second $2\pi$ is due to $\partial a$, and the last factor of $2$ is due to the equation of motion connecting $\partial a$ to $\partial A$. In result, we conclude that $\exp(i\int L_{\rm surf}) = \pm 1$ as $C = 0, \frac{1}{8\pi} mod \frac{1}{4\pi}$, and the topologically non-trivial insulator corresponds to $C = \pm \frac{1}{8\pi}$.  

   
Now, we integrate out $a$ and $b$ fields in the equation \eqref{12} with $C = \frac{1}{8\pi}$ so that we get the effective action for the electromagnetic field $A$. When we do this integration, we obtain the magnetoelectric effect of the strong topological insulator~\cite{qilong},  
\begin{equation}
S_{3D} = \pm \int d^{3}x dt \frac{1}{8\pi} \varepsilon^{\mu\nu\lambda\rho} \partial_{\mu} A_{\nu} \partial_{\lambda} A_{\rho}
\label{13}
\end{equation}

Now we introduce the theta-angle $\theta = C \times 8\pi$ which is the usual ``axion electrodynamics''~\cite{wilczekaxion} of the topological insulator.  Because of the quantization of $C$, we have $\theta \sim \theta + 2\pi$.  We allow $d\theta \neq 0$ on the surface of the topological insulator, which then influences the rest of the material through the $\partial a \partial A$ term.

With this knowledge and the full Lagrangian, we can now study other properties of the system by the equations of motion resulting from equation \eqref{12}. Before getting into the details of the equations of motion for $a$ and $b$, it is important to notice that electromagnetic current is conveyed by both $a$ and $b$. The current can be directly read off from the equation \eqref{12} by $\frac{\delta}{\delta A}L_{BF}$. 
\begin{equation}
J^{\mu}_{EM} = J^{\mu}_{b} + J^{\mu}_{a} = \frac{1}{2\pi}\varepsilon^{\mu\nu\lambda\rho} \partial_{\nu}b_{\lambda\rho} + \frac{1}{8\pi^{2}}\varepsilon^{\mu\nu\lambda\rho}\partial_{\nu}(\theta \partial_{\lambda}a_{\rho})
\label{14}
\end{equation}
So the total electromagnetic current is carried by both gauge fields $a$ and $b$. Even though the two terms of this current appear similar, the interpretation of $a$ and $b$ will show that the two are independent contribution to $J_{EM}$.

It is easy to interpret the contribution from $b$ : $b$ is a two-form and thus can be thought of as related to ordinary particle current in the same way as the field strength in electromagnetism.  Because $J_{b}$ is a free-particle current in the TIs, we might expect $J_{b}$ to appear only on the edge as the Hall current when we solve the equation of motion for $b$, and it turns out to be so. 
For $J_{a}$, we need to be somewhat careful. The term including $J_{a}$ is originally from the surface $L_{\rm surf}$. Here we treat the term as a part of bulk theory. Thus, if we only ask about the bulk current, $J_{a}=0$ trivially. But $J_{a}$ need not be zero on the surface. Actually, $J_{a} = \widehat{n} \cdot P + \widehat{n} \times M$ where $P$ is electric polarization, $M$ is magnetic polarization and $\widehat{n}$ is the normal vector to the normal. To see this, we revisit $L_{\rm surf}$ for the case $\theta=\pi$:
\begin{equation}
L_{\rm surf} = \frac{1}{8\pi} \varepsilon^{\mu\nu\lambda\rho} \partial_{\mu} a_{\nu} \partial_{\lambda} A_{\rho} = \frac{1}{2} P^{\mu\nu}\partial_{\mu} A_{\nu} = \frac{1}{2}(\overrightarrow{P}\cdot\overrightarrow{E}+\overrightarrow{M}\cdot\overrightarrow{B})
\end{equation}
Here $P^{\mu\nu}$ is the antisymmetric polarization tensor, $\overrightarrow{P},\overrightarrow{M}$ are usual electric, magnetic moments. We identify $\overrightarrow{P}^{i} = P^{0i}$ and $\varepsilon^{ijk} \overrightarrow{M}^{i} = P^{jk}$ where $P^{\mu\nu} = \frac{1}{4\pi}\varepsilon^{\mu\nu\lambda\rho} \partial_{\lambda} a_{\rho}$. The polarization has an additional constraint like usual polarization in neutral matter: $\partial_{\nu}P^{\mu\nu}=0$ implying that $\nabla \cdot \overrightarrow{P}=0 $ and $\nabla \times\overrightarrow{M}=0$. From this constraint, we can see that we split the electromagnetic response into two pieces that cannot mix up. One part is $J_{a}$ which is particle-like response, and the second part is $P^{\mu\nu}$ which is like string or polarization response. With these in mind, we can cast our $BF$ theory into the physically straightforward form 
\begin{equation}
L_{BF} = \frac{1}{2\pi} \varepsilon^{\mu\nu\lambda\rho}a_{\mu}\partial_{\nu}b_{\lambda\rho} + J_{b}^{\mu}A_{\mu} + \frac{1}{2}(\overrightarrow{P}\cdot\overrightarrow{E}+\overrightarrow{M}\cdot\overrightarrow{B})
\end{equation} 
The second term is the usual coupling of the EM field and charge current. The third term is the coupling allowed by symmetry between (magnetic and electric) polarizations and the EM field. The term shows up only in the nontrivial topological insulator. Finally, the first term gives nontrivial braiding between particle and polarizations.  This description will give in the next section the axion electrodynamics form of EM responses of topological insulators with gapped surfaces that break $T$-symmetry once we integrate out the $a$ and $b$ fields.  If on the other hand we did not break $T$- symmetry on the surfaces, then we will obtain a different description as seen below.

\begin{figure}
\includegraphics[width=0.75 \columnwidth]{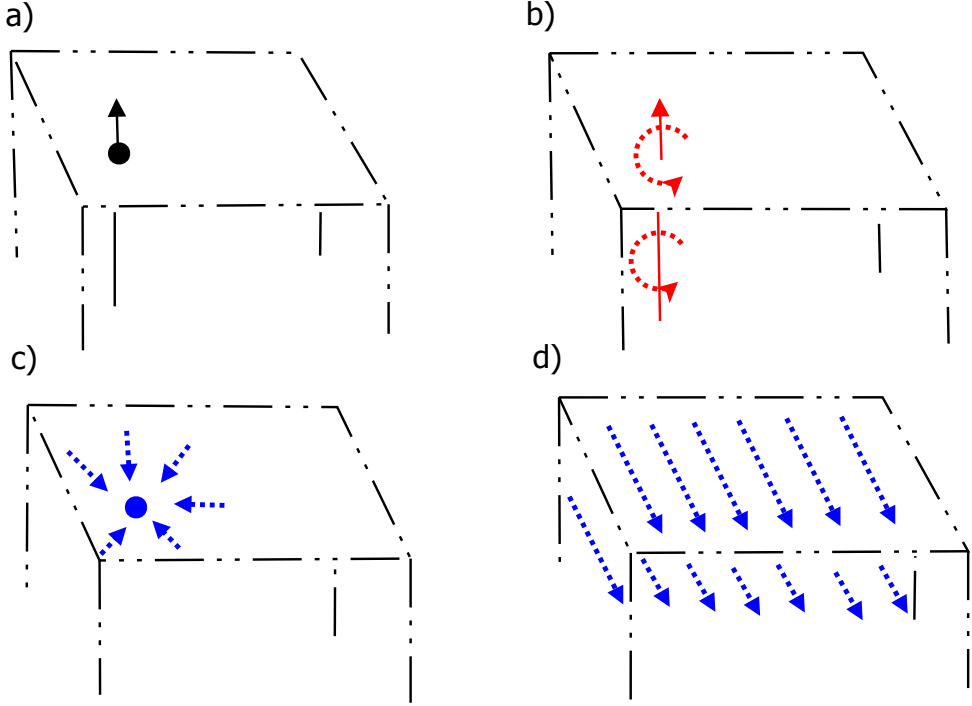}
\caption{A flux tube threads a 3D TI and penetrates the surface. a) The black arrow represents the magnetic flux and the black sphere represents the localized EM charge due to the QHE on the surface. The equations of motion for $a_{\mu}$ and $b_{\mu\nu}$ give the distribution of the gauge fields. b) The picture for $a_{\mu}$ is simple: the $a_{\mu}$ gauge field generates another flux tube at the location of the $A$ flux tube. The tensor gauge field $b_{\mu\nu}$ can be broken up into two pieces $b^{0i}$ and $\epsilon_{ijk}b^{jk}$. With the flux configuration of $A$, c) $b^{0i}$ sees a monopole source at the surface and d) $\epsilon_{ijk}b^{jk}$ sees no source but should be curl-free according to the equation of motion. The picture of $b_{\mu\nu}$ is determined only up to the gauge freedom of the field, which consists of adding any divergence-free vector field for $b^{0i}$ and adding any curl-free vector field for $\epsilon_{ijk}b^{jk}$.}
\label{Fig2}
\end{figure}

Before finishing this section, we present the equations of motion for the fields $a$ and $b$ for future use. We vary $a$ and $b$ for the Lagrangian \eqref{12}. The resulting equations are as follows :
\begin{equation} 
\frac{1}{2\pi} \varepsilon^{\mu\nu\lambda\rho}\partial_{\nu}b_{\lambda\rho} = \pm\frac{1}{8\pi^{2}}\varepsilon^{\mu\nu\lambda\rho} \partial_{\nu}\theta\partial_{\lambda}A_{\rho}, \quad \frac{1}{2\pi} \varepsilon^{\mu\nu\lambda\rho}\partial_{\lambda}a_{\rho}= - \frac{1}{2\pi} \varepsilon^{\mu\nu\lambda\rho}\partial_{\lambda} A_{\rho}
\end{equation}
Given $A$, we can obtain part of $a$ and $b$ according to these equations of motion. Effectively, $A$ is the source for $a$ and $b$. The most studied example for 3D TI is an EM flux tube threading the bulk and its surface. The gauge configuration for $a$ is easy to capture as it describes another vortex tied to EM vortex with the same magnitude. On the other hand, we need to be careful for $b$ as $b$ has more gauge degrees of freedom. For convenience, we think of a TI with $z<0$ and the magnetic flux threading the surface at the origin. We stick to the static configuration of $A$ fields. Then the equation for the gauge $b$ can be split into two parts:
\begin{equation}
J_{b}^{0} = \frac{1}{8\pi} \delta(x)\delta(y)\delta(z) = \frac{1}{2\pi} \partial_{i} \varepsilon^{ijk} b_{jk}, \quad J_{b}^{i} = 0 = \frac{1}{2\pi} (\nabla \times b^{0k})^{i} ,
\end{equation}
where we used $\partial_{0} b = 0$ for the second equation. These two equations imply that the end of the flux tube is a monopole source for $\varepsilon b_{jk}$, and $b^{0i}$ has a curl-free configuration. This is illustrated in Fig.~\ref{Fig2}.

\subsection{Gapped edge theory of 3D topological insulators}
In this section, we will establish the ``edge theory'' of the surface of a 3D TI once time-reversal-breaking perturbations have been added.  (We use edge theory in this section for comparison to the well-known edge theories of 2D topological states.  That the surface in our $BF$ theory should be $T$- broken is visible in $L_{surf}$.)  Generically the surface is gapped for $T$- symmetry-breaking perturbations present on the surface. The gapped surface should not change the bulk properties much for weak perturbations, and we expect the edge theory to reflect some properties of the bulk theory. In fact, 3D $BF$ theory generates the topological term of 2D $BF$ theory as its edge theory, plus additional terms reflecting the broken time-reversal symmetry.  The edge 2D $BF$ theory inherits topological properties from 3D bulk $BF$ theory. In this section, we derive the gapped edge theory of 3D TI in the same way as CS theory predicts a gapless edge theory of quantum Hall states, and we explore the properties of the theory including the effects of an infinitely thin flux tube threading a 3D TIs.


   

\subsubsection{Review of the edge theory of CS theory}

In this section, we briefly review the derivation of the edge theory for CS theory in 2D.  For simplicity, we ignore the coefficients and confine ourselves to understanding the requirements imposed by the gauge invariance of the theory. We follow Wen's argument leading to the edge theory of fractional quantum Hall states from gauge invariance of the theory on a manifold with boundary~\cite{wenrev1}.  Let us denote the compact bulk as $M$ and the boundary as $\partial M$. The bulk action for CS theory is $S_{0}=\int_{M} a\wedge da$ which is not gauge invariant. For a general gauge transformation $a\rightarrow a+d\gamma$, the action changes $S_{0}\rightarrow S_{0}-\int_{\partial M} d\gamma \wedge a$.

Hence we introduce a scalar field $\phi$ living on the edge and gauge transform $\phi \rightarrow \phi + \gamma$ when $a\rightarrow a+d\gamma$.  Now gauge invariance forces us to write the action for our system as $S=S_{0} - \int_{\partial M} d\phi \wedge a$. However, the Hamiltonian related to $\int_{\partial M} d\phi \wedge a$ is zero.  So we need an extra term to get the usual chiral edge theory; this term reflects physics localized near the edge, while the zero-energy Hamiltonian reflects a topological kinetic energy part coming from the bulk.

To introduce the edge term, we define the covariant derivative $D\phi = d\phi - a$, which is gauge invariant under any gauge transformation, and we write the Lagrangian $S=S_{0} - \int_{\partial M} d\phi \wedge a - C\int_{\partial M} D\phi \wedge *(D\phi)$ where $*$ is the dual operation. The key points in deriving the edge theory can be summarized as: 1) we can restore gauge invariance by introducing fields at the edge, and 2) the term $D\phi \wedge *(D\phi)$ is not coming from the gauge invariance of the bulk theory, but rather is an extra piece required for the theory to have non-trivial dynamics at the edge. This surface dependence is manifest in that the coefficient for this term in the edge chiral theory depends on the material; it determines the propagation velocity of the edge excitations. Now we use similar arguments to derive the edge theory of 3D $BF$ theory.

\subsubsection{Gapped edge theory of 3D $BF$ theory}

In this section, we derive the edge theory~\cite{EdgeBF,BalaBFedge1,BalaBFedge2} of 3D $BF$ theory by requiring the total theory with surface to have the same gauge invariance as the interior of the 3D sample. While doing so, another important feature of the bulk $BF$ term shows up; the bulk $BF$ theory is not gauge invariant, just as above for CS theory. The surplus gauge degree of freedom goes to the edge and establishes the edge theory.  We will initially study the case when the surface is gapped by a time-reversal-breaking perturbation, then return in the last part of this chapter to the question of how massless fermions appear when time-reversal symmetry is present.
As before, we ignore the numerical coefficients at first but will restore the coefficients later as the coefficients determine the statistics and Hall effect on the edge. We start with the edge theory without external EM gauge field $A$. The answer for this system can be written simply by introducing a new scalar field $\Phi$ whose values only are significant on the boundary and writing $a^\prime = a+d\Phi$: 
\begin{equation}
S = \int_{M} b\wedge da^\prime -\int_{\partial M} b\wedge a^\prime.
\end{equation}
This combined action is explicitly gauge-invariant if we define $\Phi$ to transform $\Phi \rightarrow \Phi - \gamma$ when $a \rightarrow a +d\gamma$.  (Note : $\Phi$ describes degrees of freedom only on the edge since in the bulk term $da=da^\prime$). It is elementary to check the gauge invariance when $b \rightarrow b+d\xi$. (We do not allow the variation with respect to $b$ on the edge; rather we vary $a$ and $\Phi$ independently.)  When we vary the nondynamical field $\Phi$ to obtain a constraint, we have $db=0$, i.e., $b$ is pure gauge on the edge. Thus $b=d\zeta$ at least locally for the 2D edge by Poincare's Lemma and the edge action becomes simply 2D $BF$ theory $\int_{\partial M} a\wedge d\zeta$.

When we include the coupling to an external gauge field $A$, still the same idea applies and thus we can derive the edge theory. If we add $b \wedge dA$ and $\theta da \wedge dA$ to the bulk action, then the term $b \wedge dA$ requires a term $b \wedge A$ on the edge. However, $\theta da \wedge dA$ is itself gauge invariant, so we do not need any additional term in the edge action and we can move the term $\theta da \wedge dA = d(\theta a \wedge dA)$ freely to the edge for the case $d\theta =0$ in the bulk, or vice versa. So we write the action for $M+\partial M$ as 
\begin{equation}
S = \int_{M} \left( b\wedge da' + b \wedge dA + \theta da \wedge dA \right) -\int_{\partial M} \left( b\wedge a' + b \wedge A \right)
\label{a}
\end{equation} 
We can move the term $\theta da \wedge dA$ into the edge by Gauss's Law : 
\begin{equation}
S = \int_{M} \left( b\wedge da' + b \wedge dA \right)  -\int_{\partial M} \left( b\wedge a' + b \wedge A - \theta a' \wedge dA \right)
 \label{b}
\end{equation}
The two actions (\ref{a}) and (\ref{b}) are physically equivalent. If we integrate out $a$ and $b$ in equation \eqref{a}, then we obtain $S \sim \int_{M} \theta dA \wedge dA$. On the other hand, if we integrate out $a$ and $b$ in equation \eqref{b}, we obtain $S \sim \int_{\partial M} A \wedge dA$, identical to the case of equation \eqref{a}. This illustrates that we can move every physical response to the EM gauge field to the edge by moving the term $\theta da \wedge dA$. The effect of pushing this term to the edge is not only a surface Hall effect but also statistical, and we will discuss this effect again later. For later use we restore the important coefficients for the action now. We take the edge to lie in the $xy$ plane and assume that $b$ is a the pure gauge $d\zeta$ on the edge.  Then
\begin{eqnarray}
S &=& \int_{M} \frac{1}{2\pi} \varepsilon^{\mu\nu\lambda\rho}b_{\mu\nu}\partial_{\lambda}a_{\rho} +\frac{1}{2\pi} \varepsilon^{\mu\nu\lambda\rho}b_{\mu\nu}\partial_{\lambda}A_{\rho} + \frac{\theta}{8\pi^{2}}\varepsilon^{\mu\nu\lambda\rho} \partial_{\mu}a_{\nu}\partial_{\lambda}A_{\rho} \cr
&&- \int_{\partial M}\frac{1}{\pi}\varepsilon^{\mu\nu\rho} \zeta_{\mu}\partial_{\nu} a_{\rho} + \frac{1}{\pi}\varepsilon^{\mu\nu\rho} \zeta_{\mu}\partial_{\nu} A_{\rho}
\end{eqnarray}
Some comments are worthwhile here. First, the gauge invariance of the theory is guaranteed by $\zeta_{\mu} \rightarrow \zeta_{\mu} + \xi_{\mu}$ when $b_{\mu\nu} \rightarrow b_{\mu\nu} + \partial_{\mu}\xi_{\nu} -\partial_{\nu}\xi_{\mu}$, and the gauge transformation of $a$ leaves the theory invariant as $a$ appears only as $d\wedge a$.

Secondly, when we move the term $\theta da \wedge dA$ into the edge as $\theta a \wedge dA$, $\theta$ term appearing in the edge action should be interpreted as $\delta \theta = \theta_{TI}-\theta_{vacuum}$, and $\theta$ itself is not well-defined on the edge so the term proportional to $da\wedge dA$ is breaking $T$ symmetry. Lastly, the sources of $b$ in bulk and $\zeta$ on the surface are the same. If there is a source $\tau$ of $\zeta$ on the surface that enters as $\tau^{\mu}\zeta_{\mu}$, then $\tau$ should be a part of $\Sigma_{\mu\nu}$ threading the surface (explicitly, $\tau^{0} = \widehat{n}_{i} \cdot \Sigma^{0i}$ where $\widehat{n}$ is the normal vector to the surface). Thus, there is no local excitation at the surface for $b$ or $\zeta$ fields. With the full machinery, we take the edge action of the 3D bulk action as the following : 
\begin{equation}
S_{edge}= \int_{\partial M}\frac{1}{\pi}\varepsilon^{\mu\nu\rho} \zeta_{\mu}\partial_{\nu} a_{\rho} + \frac{1}{\pi}\varepsilon^{\mu\nu\rho} \zeta_{\mu}\partial_{\nu} A_{\rho} +\frac{\theta}{8\pi^{2}}\varepsilon^{\mu\nu\rho} a_{\mu}\partial_{\nu}A_{\rho}
\label{c}
\end{equation}
For the bulk action of equation \eqref{c}, we have 
\begin{equation}
S = \int_{M} \frac{1}{2\pi} \varepsilon^{\mu\nu\lambda\rho}b_{\mu\nu}\partial_{\lambda}a_{\rho} +\frac{1}{2\pi} \varepsilon^{\mu\nu\lambda\rho}b_{\mu\nu}\partial_{\lambda}A_{\rho},
\end{equation}
which is trivial for the $A$ field when $a$ and $b$ are integrated out: in this form the entire electromagnetic response comes from the surface.

\subsubsection{Infinitely thin flux tube}

In this section, we calculate the charge bound to a flux tube by using the gapped edge action we derived in the previous section.  This provides some intuition for the meaning of the fields in the effective theory.  We thread a $\pi$ flux tube through the 3D TI adiabatically. The flux tube is inserted into a thin cylinder in 3D TIs, and the surface of the cylinder is itself an edge of the 3D TI. While doing so, we will use the gapped edge action because the cylinder is also part of the edge and thus gapped (i.e., we assume the same $T$-breaking perturbation continues along the cylinder). Unfortunately, we cannot get the gapless edge theory discussed in the next section from the gapped edge action by naively plugging in a $\pi$ flux configuration of $A$ field, but we can glimpse the effects of degrees of freedom living in the $\pi$ flux tube of 3D TIs.

From the action for edge \eqref{c}, there can be at most two degrees of freedom carried by $\zeta$ and $a$. Thus, we can guess there are possibly two chiral modes to the gapless cylinder surface because of time reversal properties(or no localized mode); indeed there are two gapless chiral modes at the infinitely thin cylinder with $\pi$ flux as expected.  Now, we calculate the charge bound to the $\pi$ flux at the end of the flux tube. For simplicity, we imagine an infinite bulk with the cylindrical hole at the center and assume that the flux remains spatially separated from the surface of the cylinder. We take the $x$-axis along the cylinder axis and the $y$-axis around the cylinder on the surface(We compactify $y$ direction). Thus, the flux $\Phi$ threaded in the cylinder is given by the integral of $A_{y}$ along the $y$ direction.  
\begin{equation}
\Phi(t) = \oint A_{y}(t) dy
\end{equation}
Given this external field $A$, we have equations of motion for $\zeta$ and $a$. Before deriving the equations of motion, we first identify the EM current due to external EM gauge field by $\frac{\delta}{\delta A} S$ where $S$ is the action for the edge: 
\begin{equation}
J_{EM}^{\mu} = j_{\zeta}^{\mu}+j_{a}^{\mu}= \frac{1}{\pi}\varepsilon^{\mu\nu\rho} \partial_{\nu}\zeta_{\rho} +\frac{1}{8\pi}\varepsilon^{\mu\nu\rho}\partial_{\nu}a_{\rho}
\end{equation} 
where we plugged $\theta=\pi$. The equations of motion for $\zeta$ and $a$ are easily derived by varying with respect to $a$ and $\zeta$. 
\begin{equation}
j_{\zeta}^{\mu} = \frac{1}{8\pi}\varepsilon^{\mu\nu\rho} \partial_{\nu}A_{\rho}, \qquad
j_{a}^{\mu} = \frac{1}{8\pi}\varepsilon^{\mu\nu\rho} \partial_{\nu}A_{\rho}
\end{equation}
Thus, we have the following equation for the total current flowed along the cylinder. 
\begin{equation}
Q = \int dt J_{EM}^{x} = \int\int dtdy \frac{1}{4\pi} \partial_{t}A_{y} = \frac{\Phi}{4\pi}
\end{equation}
Because the total flux at the end of threading process is $\pi$, we obtain charge $\frac{e}{4}$ localized at the one side end of the flux tube and $-\frac{e}{4}$ on the other side end of the flux. We wish to emphasize that this localized charge is exactly what is expected from the surface hall effect. When the surface of the TI is gapped, the surface hosts $\frac{1}{8\pi}$(half the conductance quantum; $e = \hbar = 1$) Hall conductance, and thus we have charge $\frac{e}{4}$ when $\pi$ flux is threaded. One more comment is in order : as the surface explicitly breaks $T$- invariance, we can deduce that the system distinguishes $\pi$ from $-\pi$ as far as the flux threads the surface. A consequence is that the charge localized at the tip of $\pi$ flux is different from that of the $-\pi$ flux. However, if the flux totally lies in the bulk by making it a torus, then the system never distinguishes $\pi$ flux from $-\pi$ flux. This can be easily seen from the bulk Lagrangian. 
\begin{equation}
L = \frac{1}{2\pi} \varepsilon^{\mu\nu\lambda\rho}a_{\mu}\partial_{\nu}b_{\lambda\rho} +\frac{1}{2\pi} \varepsilon^{\mu\nu\lambda\rho}A_{\mu}\partial_{\nu}b_{\lambda\rho}
\end{equation}
From this Lagrangian, we see that $\pi$ flux is one unit of vorticity of the $b$ gauge field. Upon encircling a unit source of $a$ around the $\pi$ flux, the wave function obtains a phase factor $\exp(i\pi)$ which is equivalent to $\exp(-i\pi)$ where $-\pi$ flux is in place of $\pi$ flux.  Having reproduced the expected properties for gapped surfaces, we now turn to the emergence of Dirac fermions at the surface of $BF$ theory when time reversal is unbroken.

\subsubsection{Gapless edge theory of 3D $BF$ theory}

Rather than a problem, it is a useful feature of 3D $BF$ theory that the bulk theory by itself is not invariant under gauge transformations on a manifold with boundary.  The gauge fields forced to live at the edge cancel out the surplus gauge freedom of the bulk theory.  Here we will study the gapless surface of 3D $BF$ theory and will find explicit expressions for a surface Dirac fermion field, constructed from the bosonic fields topologically required to exist at the surface of a 3D topological insulator. Combined with the ability to reproduce the expected electromagnetic coupling for gapped surfaces, this fermionic boundary suggests strongly that 3D $BF$ theory is the proper topological field theory for the 3D strong topological insulator. 

The basic idea is that 3D $BF$ theory contains two bosonic fields at the edge~\cite{HanssonBF}, one arising from $a_{\mu}$ and the other from $b_{\mu\nu}$. From these two bosonic fields, it is not obvious at first sight how to construct a fermion. Fortunately, there are bosonized theories for (2+1)D fermions (and other higher dimensions as well~\cite{LutherBosonization}) which were originally constructed to generalize the well-known bosonization of (1+1)D fermionic systems.  The key idea of these constructions, which can be made precise in terms of the ``topographic'' representation~\cite{Bosonization3,Bosonization1}, lies in describing the higher-dimensional system as an infinite number of (1+1) dimensional 'rays'.  The complicated part is in maintaining the Fermi statistics of particles belonging to different rays.  We will find that the Dirac fermion constructed out of the two bosons predicted by $BF$ theory shows the special property of the surface of 3D topological insulators that the momentum and the spin of electrons are locked to each other.



In this section, we first start with the edges of 2D $BF$ theory to clarify the structure of the surface edges of 3D $BF$ theory, and move into the surface of 3D $BF$ to obtain a Dirac fermion. While we are dealing with the surface of 3D $BF$ theory, we show that there are two gapless bosonic fields on the surface of 3D $BF$ theory and proceed to the ``tomographic'' transformation of introducing rays.  We can then compare the result to the bosonization of a single Dirac fermion.

In order to emphasize the similarities between the 2D and 3D TIs in the $BF$ description, we review the edge of the 2D $BF$ theory first and see how the surface theory of 3D $BF$ can be thought of as an infinite number of ``rays'' described by the edge theory of 2D $BF$ theory. To get the edge theory of 2D $BF$ theory, we start by solving the bulk equations of motion of the 2D bulk $BF$ theory, i.e $a_{k} = \partial_{k} \Lambda$, and $b_{k} = \partial_{k} \Gamma$ with Coulomb gauge $a_{0}=b_{0}=0$.  Up to a constant,
\begin{equation}
\int_{M} L = \int_{M} \varepsilon^{\mu\nu\lambda}a_{\mu} \partial_{\nu} b_{\lambda}  = \int_{\partial M}(\partial_{t} \Gamma) (\partial_{x} \Lambda) + (\partial_{x} \Gamma) (\partial_{t} \Lambda)
\label{T1}
\end{equation}
where $M$ is 2D bulk, and the dynamics on its edge $\partial M$  is described by the fluctuations of $\Lambda$ and $\Gamma$.  We see that there is a first-order edge Lagrangian from the bulk: the Hamiltonian is zero until we add additional non-universal terms generated by the boundary.

The physics becomes more transparent if we take symmetric and antisymmetric combinations of $\Lambda$ and $\Gamma$.  Substituting $S = (\Lambda + \Gamma)/2$ and $A = (\Lambda - \Gamma)/2$ into the equation (\ref{T1}), 
\begin{equation}
\int_{\partial M}(\partial_{t} \Gamma) (\partial_{x} \Lambda) + (\partial_{x} \Gamma) (\partial_{t} \Lambda) = \int_{\partial M}(\partial_{t} S) (\partial_{x} S) - (\partial_{x} A) (\partial_{t} A)
\end{equation}
The surface is assumed to generate the term
\begin{equation}
H = \int \frac{v_1}{2} (\partial_{x}A)^{2} + \frac{v_2}{2} (\partial_{x}S)^{2}
\label{T2}
\end{equation}
Where $v_1$ and $v_2$ are the speeds of fluids and is system dependent. We identify the fluid density operators $\rho_{1}(x) = \partial_{x} S$ and $\rho_{2}(x) = \partial_{x} A$. The equations of motions for $A$ and $S$ are now understood as the continuity equations for the fluids,
\begin{equation}
\partial_{t} \rho_{i}(x,t) + (-1)^{i}v \partial_{x} \rho_{i}(x,t) = 0,\quad i = 1,2
\end{equation}
Thus, there are two chiral bosonic modes flowing in opposite directions. Moreover, we can see that the spectrum of the Hamiltonian (\ref{T2}) is $v|k|$ after quantization where $|k|$ is the magnitude of the one-dimensional momentum, and time-reversal symmetry forces $v_1=v_2$.  These bosonic modes can be viewed as the density excitations of a one-dimensional fermionic system via standard 1+1-dimensional bosonization.

We now wish to follow the same general prescription for the surface of 3D $BF$ theory: we obtain bosonic fields and a kinetic term from the bulk action and add potential terms assumed to arise from non-universal physics at the surface.  The result is then shown to be a bosonized version of a 2+1-dimensional free Dirac fermion satisfying the expected conditions for a TI surface.
Starting with the $BF$ theory with no $T$-breaking surface term and solving the bulk equations of motions for $a$ and $b$ with Coulomb gauge ($a_{i} = \partial_{i}\Lambda, b_{ij} = \partial_{i}\wedge \zeta_{j}$, and $a_{0} = b_{0i}=0$), we obtain (up to an overall constant)
\begin{equation}
\int_{M} L = \int_{\partial M} \partial_{t}\Lambda \varepsilon^{ij}\partial_{i}\zeta_{j} + \partial_{i}\Lambda \varepsilon^{ij}\partial_{t}\zeta_{j}
\label{S1}
\end{equation}  
Hence the surface theory contains a scalar field $\Lambda$ and a vector field $\zeta$, from which we need to construct fermions.  The same surface theory for 3D $BF$ theory has been obtained previously~\cite{HanssonBF}; we are not aware that the following connection between $BF$ theory and surface Dirac fermions or topological insulators has been made before.  The topological part of the surface action for $BF$ theories more generally is discussed in Appendix B of Ref.~\citenum{moorebf}.

We assume that the effect of non-universal surface physics is to generate the potential terms
\begin{equation}
H = \int (\alpha_1 (\partial_{i}\Lambda)^{2} + \alpha_2 (\varepsilon^{ij}\partial_{i}\zeta_{j})^{2})
\label{S2}
\end{equation}
analogous to the potential terms at the quantum Hall edge.  The constants $\alpha_1$ and $\alpha_2$ are non-universal and will ultimately determine the velocities of excitations, which we assume to be isotropic; henceforth we set $\alpha_1 = \alpha_2 = 1$ so that the velocities of propagation are unity and we can connect to some previous field theory literature.  The surface equation of motion from varying $\Lambda$ is
\beq
\partial_t \varepsilon^{ij}\partial_{i}\zeta_{j} - \nabla^2 \Lambda = 0
\eeq
which is the total derivative of
\beq
\partial_t \varepsilon^{ij} \zeta_{j} - \partial_i \Lambda = 0.
\label{eqm1}
\eeq
The equation of motion from varying $\zeta_i$ is similarly the total derivative of
\beq
\partial_t \Lambda - \partial_i \varepsilon^{ij} \zeta_j = 0.
\label{eqm2}
\eeq
But (\ref{eqm1}) and (\ref{eqm2}) are precisely the ``Bose structure'' on one scalar and one vector boson that are required to construct massless fermionic fields via the tomographic representation; the key equation (21) of Ref.~\citenum{Bosonization1} is just the above with $B \rightarrow \Lambda$, $A \rightarrow \zeta$, and we switch to that paper's notation in the following.  For self-containedness and to make this result less mysterious, we reproduce the key steps from that work and the explicit expression for the Dirac spinors.  Note that the first-order kinetic energy term obtained from $BF$ theory is not what would be obtained from, e.g., fluctuations of an elastic surface; a bosonization approach to the TI surface seemingly inequivalent to ours has recently been proposed by Vildanov~\cite{BosonizationTI} starting from a modified hydrodynamics of an incompressible fluid.

We now proceed to the tomographic representation for the scalar $A$ and vector $B$., and we will quantize them to obtain two bosonic fields describing two incompressible fluids. The tomographic representation uses the following generalized delta function:
\begin{equation}
\delta^{1/2}(y-\hat{n}\cdot {\bf r}) = \frac{1}{2\pi} \int_{-\infty}^{\infty} dk |k|^{1/2} e^{iky}e^{-ik\hat{n}\cdot {\bf r}} = \frac{1}{2\pi}\int_{0}^{\infty} dk k^{1/2} \cos k(y-\hat{n}\cdot {\bf r})
\end{equation}
This function is used to transform the usual cartesian representations of fields into the tomographic representation. For example, the tomographic transformation of a scalar field $\phi({\bf r})$ is defined as 
\begin{equation}
{\tilde \phi}(y,\hat{n}) = \int d{\bf r} \delta(y-\hat{n}\cdot {\bf r}) \phi({\bf r}).
\end{equation}
Physically, $y\in(-\infty,\infty)$ represents the projected distance of ${\bf r}$ from the origin along the direction $\hat{n}$. From now on, we write $\theta$  for the direction of $\hat{n}$, i.e $\hat{n} = (\cos(\theta), \sin(\theta))$.  Scalar, spinor, and vector fields in the tomographic coordinates are denoted with an extra tilde.

From this definition, the following two useful relations can be obtained:
\begin{equation}
\delta(r-r') = \frac{1}{4\pi} \int dy d\theta \delta(y-\hat{n}\cdot r) \delta(y-\hat{n}\cdot r'),
\end{equation}
\begin{equation}
\frac{1}{2\pi} \int d^{2}r \delta(y-\hat{n}\cdot r) \delta(y'-\hat{n'}\cdot r) = \delta(y,y')\delta(n,n') + \delta(y,-y')\delta(n,-n')
\label{EQ11}
\end{equation}
Note that the second equation manifests the connection between $(\hat{n},y)$ and $(-\hat{n},-y)$ that appears below in the identification of the surface theory as a single Dirac fermion in condensed matter language.  These are the standard tools for the tomographic representations of bosonization theory~\cite{Bosonization1,Bosonization2,Bosonization3}.

With the radiation gauge for $A$ ($\nabla \cdot A = 0$), after tomographic transformation of $A$ and $B$ one obtains~\cite{Bosonization1}
\beqn
\partial_0 {\tilde B}(y,{\hat n}) &=& \partial_y^T {\tilde A}(y,{\hat n}) \cr
\partial_y {\tilde B}(y,{\hat n}) &=& \partial_0^T {\tilde A}(y,{\hat n}),
\eeqn
and the longitudinal part of ${\tilde A}$ is constant and can be ignored.  These describe bosons propagating with nonzero velocity along each ray direction ${\hat n}$.  A spinor field in tomographic coordinates can be constructed for each ray as a normal-ordered version of
\beq
{\tilde \psi}(y,{\hat n}) = C \exp(i \sqrt{\pi} \left[{\tilde A}(y,{\hat n}) + {\tilde B}(y,{\hat n})\right],
\eeq
where $C$ is a normalization constant.  The last step is to add a 2D ``Klein factor'' that ensures the canonical anticommutation relations of fermions are satisfied not just on one ray but between different rays; the end result is that $\tilde \psi$ is multiplied by an operator~\cite{Bosonization1}
\beq
O_{\hat n} = \exp\left( {i \sqrt{\pi} \over 2} \int_0^\theta\,d\theta^\prime\,\left[ \alpha(\hat n(\theta^\prime)) + \beta(\hat n(\theta^\prime))\right]\right)
\eeq
where the charges $\alpha$ and $\beta$ are
\beqn
\alpha({\hat n}) &=& \int_{-\infty}^\infty \partial_0 {\tilde A}(y,{\hat n}) \cr
\beta({\hat n}) &=& \int_{-\infty}^\infty \partial_0 {\tilde B}(y,{\hat n}).
\eeqn
This fermion propagates with unit velocity in the $+y$ direction and is the tomographic transform of a 2D massless Dirac fermion.  A subtle point is that there is another fermion field~\cite{Bosonization1} ${\tilde \chi}$ propagating in the $-y$ direction, but we note that the Hermitian conjugate of ${\tilde \chi}(-y,-{\hat n})$ is equivalent to ${\tilde \psi}(y,{\hat n})$ up to a phase factor and conclude that the above indeed describes just one fermion propagating in each direction; note that the surface-dependent chemical potential needs to be specified to determine the relative excitation energies of ``electron'' and ``hole'' propagating in a given direction.

Hence the physical picture of the gapless edge of $BF$ theory is as follows.  The scalar and vector boson fields have a kinetic term from the bulk theory and a potential term from the surface theory, as in the Chern-Simons case.  These combine to allow the faithful representation of Fermi fields.  The details of the surface determine both the velocities on each ray and the filling of the Fermi states, i.e., the chemical potential.  It is well understood~\cite{houghtonmarston,HaldaneBosonization} that the remarkable power of bosonization for interacting systems in one dimension does not carry through straightforwardly to higher dimensions, essentially because the ray decomposition of the Hilbert space (a ``superselection rule'') is less useful for normal interactions.  The $BF$ theory of topological insulators predicts that interactions in the bulk still lead to massless Dirac fermions via the bulk-edge connection derived in this section, and additional interactions occurring at the surface can be more naturally treated in the emergent fermionic variables than the bosonic ones.  A future direction that we comment on briefly below is to consider the surfaces of fractional topological insulators in three dimensions.

\begin{figure}
\includegraphics[width=0.75\columnwidth]{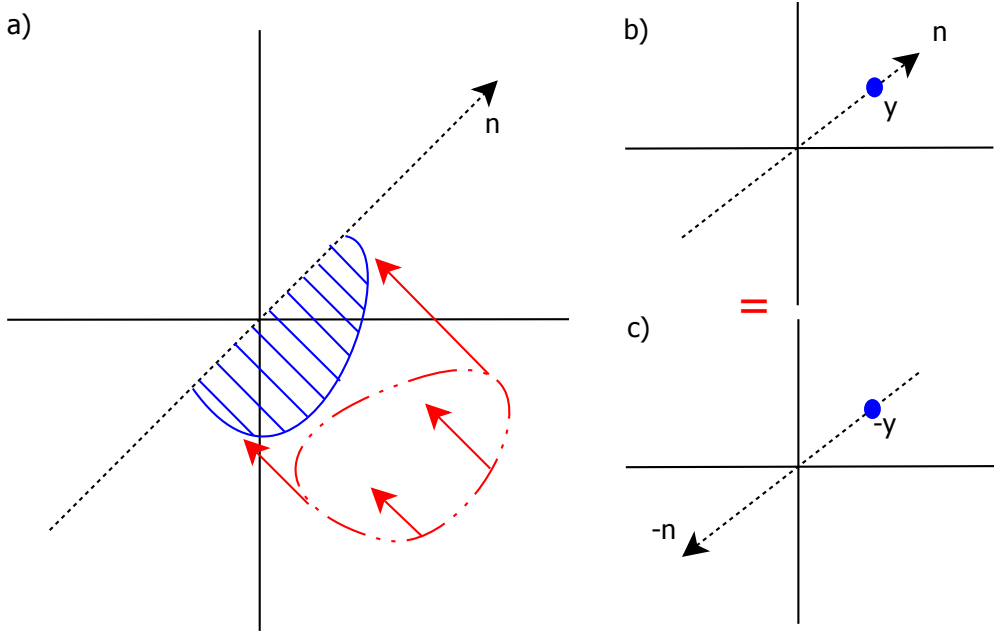}
\caption{a) Illustration of the tomographic representation of a distribution $P=P(x,y)$, which is depicted as the red object. The tomographic projection is to project the distribution on the line along $\hat{n}$ and is described as the blue dashed density on the line spanned by $\hat{n}$. b) and c) explain why $(\hat{n},y)$ is the same as $(-\hat{n},-y)$, as a projection onto $(\hat{n},y)$ is equivalent to the projection onto $(-\hat{n},-y)$.}
\label{Bos1}
\end{figure} 

\section{Conclusions and future directions}

We have argued that a topological field theory of $BF$ type is the effective theory of 2D and 3D topological insulators.  For the 2D topological insulator, this can be understood easily when one direction of spin is conserved and the system separates precisely into two copies of the quantum Hall effect.  The $BF$ description of the 3D topological insulator is our main result; perhaps the most surprising property is the emergence of surface Dirac fermions from the particular kinetic term of scalar and vector bosonic fields that are forced to exist by the incomplete gauge invariance of $BF$ theory.

There are several obvious generalizations and future directions that can be pursued.  Considerable recent interest has gone into classifying topological insulators and superconductors with other symmetries than time-reversal, including their responses with gapped surfaces~\cite{ryuresp,zhangresp}, which are the analogues of the ${\bf E} \cdot {\bf B}$ response in the conventional topological insulator.  Such systems may also be described by topological field theories, and it would be interesting to find the appropriate theories for those systems and the field-theory description of their defects and interfaces.

We will focus in closing on one particular direction where we believe $BF$ theory holds considerable promise.  Consider the Chern-Simons effective theory of the integer quantum Hall effect.  Multiplying the coefficient of the Chern-Simons term by 3 immediately gives the essential features of the {\it fractional} quantum Hall effect state previously described microscopically by Laughlin, including the braiding statistics and modified scaling dimension of the electron operator at the edge, which can be probed in tunneling experiments.

In the same way, we can obtain an effective theory for potential fractional 3D topological insulators by modifying the coefficient of the $BF$ term.  It should be stated from the outset that there is not yet a microscopic parent Hamiltonian for a 3D fractional topological insulator.  Recent constructions of effective descriptions of 3D fractional topological insulators have been based on ``parton'' ideas: the electron fractionalizes into a combination of other particles, one of which (a neutral spinon~\cite{pesinbalents} or a fractionally charged object~\cite{swingle,zhangparton}) then forms a topological insulator state.  The $BF$ theory presented here leads to a picture of 3D fractional topological insulators that appears somewhat different.  Changing the coefficient of the braiding term to another of the set of possible values~\cite{moorebf} leads to fractional statistics between point-like and line-like objects.  For a single pair of fields $(a_\mu, b_{\mu \nu})$, gapped surfaces will have surface quantum Hall layers of Hall conductance $1/2, 1/6, 1/10, \ldots$, where the first is the ``integer'' topological insulator studied in this paper and all surface Hall conductances are ambiguous by an integer.

The most interesting aspect of such a fractional topological insulator might be its gapless surface theory.  Each 1D ray in the tomographic representation is now a pair of chiral Luttinger liquids with renormalized scaling dimension of the electron operator, i.e., essentially the effective theory discussed previously~\cite{levinstern} for the edge of a 2D fractional topological insulator.  The 2D surface state formed from the collection of such rays would be an unusual non-Fermi-liquid state worthy of future study.  Leaving this fractional case aside, we can conclude from the results in this paper that $BF$ theory can describe the universal physics of the currently extant topological insulators just as Chern-Simons theory captures the universal physics of quantum Hall states.

\bigskip
\noindent\textbf{Acknowledgements}
\medskip

The authors thank O.~Ganor and T.~H.~Hansson for helpful comments.  They acknowledge support from FENA (G.~Y.~C.) and NSF DMR-0804413 (J.~E.~M.).

\bigskip
\noindent\textbf{References}


\end{document}